\documentclass[letter,superscriptaddress,twocolumn,prl,showkeys,showpacs,preprintnumbers,superscriptaddress]{revtex4-2}

	\usepackage{amsmath} 
	\usepackage{makeidx}
	\usepackage{amsfonts}
	\usepackage{graphicx}
	\usepackage[ansinew]{inputenc}
	\usepackage{subfigure}
	\usepackage{epsfig}
	\usepackage{float}\usepackage[colorlinks,hyperindex]{hyperref}
	\hypersetup
	{
		colorlinks,%
		citecolor=black,%
		linkcolor=black,%
		urlcolor=black,%
	}

	\usepackage{mathrsfs}
\usepackage{color}
\definecolor{myColor}{rgb}{0.9,0.9,0.9}

\usepackage{orcidlink}

\newtheorem{thm}{Theorem}

\newcommand{\ba}{\begin{array}}
\newcommand{\ea}{\end{array}}

\newcommand{\bc}{\begin{cases}}
\newcommand{\ec}{\end{cases}}

\newcommand{\p}[1]{\partial_{#1}}

\newcommand{\eq}[1]{\begin{equation}#1\end{equation}}
\newcommand{\alg}[1]{\begin{aligned}#1\end{aligned}}

\makeindex

\begin{document}

\title{Phase transition between shock formation and stability in cosmological fluids}

\author{David Fajman\orcidlink{0000-0003-3034-6232}}
\affiliation{Faculty of Physics, University of Vienna, 
Austria.}
\email{david.fajman@univie.ac.at}
\author{Maciej Maliborski\orcidlink{0000-0002-8621-9761}}
\affiliation{Faculty of Mathematics, University of Vienna, 
Austria.}
\author{Maximilian Ofner}
\affiliation{Faculty of Physics, University of Vienna, 
Austria.}
\affiliation{Vienna Doctoral School in Physics, University of Vienna, 
 Austria.}
\author{Todd Oliynyk\orcidlink{0000-0003-3457-9578}}
\affiliation{School of Mathematics, Monash University,
Australia.}
\author{Zoe Wyatt\orcidlink{0000-0001-5120-1839}}
\affiliation{Department of Pure Mathematics and Mathematical Statistics, 
University of Cambridge, UK.}

\date{\today}

\begin{abstract}
We demonstrate a novel phase transition from stable to unstable fluid behaviour for fluid-filled cosmological spacetimes undergoing decelerated expansion. This transition occurs when the fluid speed of sound $c_S$ exceeds a critical value relative to the expansion rate $a(t)=t^{\alpha}$ of spacetime. We present an explicit relationship between $\alpha$ and $c_S$, which subdivides  the $(\alpha,c_S)$-parameter space into two regions.  Using rigorous techniques, we establish stability of quiet fluid solutions in the stable region. Numerical experiments  reveal that the complement of the stable region consists of unstable solutions, implying sharpness of our stability result. We provide a definitive analytical bound and high-precision numerical evidence for the exact location of the critical line separating the stable from the unstable region. 

\end{abstract}


\maketitle

\section{Introduction} 
The standard model of cosmology features three key components: radiation, matter (where the pressure is much smaller than the energy density) and the cosmological constant $\Lambda$ or dark energy (cf.~\cite{Dodelson:2003ft,Durrer,Baumann}). Radiation and matter components are typically modelled as perfect fluids obeying the relativistic Euler equations
\eq{\label{Euler}\alg{
u^\mu\nabla_\mu \rho+(\rho+p)\nabla_\mu u^\mu&=0\\
(\rho+p) u^\mu \nabla_\mu u^\nu+(g^{\mu\nu}+u^\mu u^\nu)\partial_\mu p&=0.
}}
Radiation and matter are the dominant contributions to the expansion of the Universe in the early and recent stages respectively, where fluid density components lead to  expansion rates $a(t)=t^{\frac{1}{2}}$ (radiation) during the early Universe followed by a period with $ a(t)=t^{\frac{2}{3}}$ (dust) in the matter phase. Key events for the present large-scale structure of the Universe occurred during such phases and are closely connected to fluid shock formation  \cite{BiBuKa92,RKHJ03,PT16,MRKJCO00,QIS98}. The cosmological constant or dark energy is considered the main mechanism driving accelerated expansion in  later stages of cosmological evolution.

	This letter concerns the radiation and matter epochs, and their transition period, for a spatially flat universe, as consistent with the current interpretation of observations, ie.~the post-CMB epochs of evolution according to the standard model. 
While the development of shocks from significant large inhomogeneities seems natural, and is known to occur in particular cases \cite{ReSt08}, it is unknown under what circumstances shocks  form from small inhomogeneities in the fluid. We use the term \emph{fluid instability} to refer to shocks that develop from {arbitrarily small} inhomogeneities of a quiet fluid solution (ie. spatially homogeneous solution with $u^\mu \equiv (\partial_t)^\mu$).
Our results imply that for \emph{all} decelerated epochs there are unstable fluids.
\subsubsection{Fluid stabilization}
In the class of barotropic fluids with speed of sound $c_S=\sqrt{K}$ and equation of state
\eq{\label{eos}
p=K\rho,
} 
the global-in-time regularity of the fluid (in the expanding time direction) is determined by the expansion rate of the Universe, ie.~the scale factor $a(t)$, when the spacetime metric is of the form
\eq{\label{spacetimemetric}
-dt^2+a(t)^2g 
}
for a fixed Riemannian 3-metric $g$. Sufficiently fast expansion induces a damping effect in the fluid, which reduces the tendency for shocks to form. This effect was first studied in the context of Newtonian cosmology undergoing accelerated expansion \cite{BRR94}. This stabilization effect competes with the speed of sound, which, if sufficiently large, increases the tendency of shocks forming. 

This behaviour is well understood in three regimes. In the absence of expansion ($\dot a(t)=0$) quiet fluids are unstable \cite{Ch07}. In the regime of accelerated expansion ($\ddot a(t)>0$) such fluids are stable. Stability means that small inhomogeneities remain uniformly bounded and the fluid remains regular for all times. Stability has been rigorously established during the last two decades in an extensive series of works \cite{RoSp13,Sp12,Oliynyk16,HaSp15,LVK13,Fr17,W18,LFW21,M22}. Notably, in the regime of accelerated expansion even superradiative fluids ($K>1/3$) are known to be stable \cite{Oliynyk:2021,MO22,FMO24} in the tilted case, but unstable in the homogeneous untilted case \cite{BMO23,O23}.
For the case of linear expansion ($a(t)=t$), stability depends non-trivially on the speed of sound: radiation fluids are unstable \cite{Sp13} while all barotropic subradiative fluids ($K<1/3$) are stable \cite{FOW:2021,Sp13}. Stability for subradiative fluids in the regime of linear expansion holds even in the presence of backreaction \cite{FOOW:2024,FOfW:2021}. 

The critical regime, that of decelerated expansion, has so far been poorly understood  and is the subject of this letter. This regime includes the radiation and matter-dominated epochs and their intermediate epochs. 
\subsubsection{Conditions for instability}
\begin{figure}[h!]
	\includegraphics[width=\columnwidth]{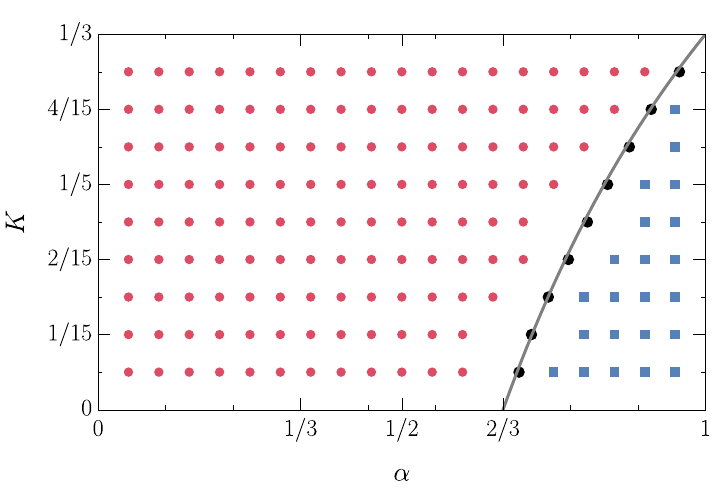}
	\caption{Numerical results for the parameter space: Stable (blue) and unstable (red) regions in the parameter space $(\alpha,K)$. The black points are determined from the asymptotic analysis of the unstable points (see the text in section \emph{Numerics on small data shock formation} below) and their location coincide with the critical curve \eqref{transitionline} (grey line). The grey curve is independently predicted by a rigorous stability analysis as a lower bound for values of $K(\alpha)$, such that $(\alpha,K)$ is unstable (see section \emph{Decay of perturbations in the stable regime}).  
		}
\label{fig:n2}
\end{figure}

We consider the full range of decelerated power-law expansion rates
\eq{
a(t)=t^{\alpha},\qquad 0<\alpha<1,\label{expansionrate}
}
and fluids with speeds of sound exhausting all speeds between dust and radiation
\eq{
0\leq K\leq 1/3.
} 
We provide strong evidence that the parameter space 
$$
\left\{(\alpha,K)\left.\right|\,\alpha\in(0,1),\,K\in[0,1/3]\right\}
$$
decomposes into two regions: a region of stable pairs and a region of unstable pairs (cf.~Fig.~\ref{fig:n2}). A pair $(\alpha,K)$ is \emph{unstable} if, in the spacetime \eqref{spacetimemetric} with $\alpha$-expansion rate \eqref{expansionrate}, there exists a sequence of initial data approximating a quiet fluid solution, that launches a sequence of solutions for which  singularities form in finite time. Our analytical and numerical studies illustrate that the transition between both regions is located on the \emph{critical line}, \eq{\label{transitionline}
K_{\mathrm{crit}}(\alpha)=1-\frac{2}{3\alpha}.
}
For an epoch with expansion rate $\alpha \in (2/3,1)$ the phase transition of a quiet fluid occurs when the speed of sound passes the critical value $K_{\mathrm{crit}}(\alpha)$. 
This yields the following primary observations:
\begin{itemize}
\item In the decelerated regime, stability of the fluid depends on the position in parameter space relative to the critical curve \eqref{transitionline}.
\item All barotropic fluids are unstable when $\alpha<2/3$ . 
\item For expansion rates $\alpha>2/3$ the barotropic fluids are stable if $K<K_{\mathrm{crit}}(\alpha)$ and unstable otherwise.
\item The behaviour in the decelerated regime is in contrast to the accelerated regime, where fluid stabilization is known to be universal. It follows that relativistic fluids present in the Universe during an epoch of decelerated expansion  must have formed shocks, provided their speed of sound was sufficiently small relative to the expansion rate. 

\item For expansion rates in the regime $1/2<\alpha\leq2/3$, dust fluids are stable. On the other hand, fluids with arbitrarily small but positive speed of sound are unstable. Thus in this regime, which interpolates  between the standard radiative and dust models, the behaviour of dust does not approximate the behaviour of fluids with non-vanishing speed of sound, and so using a dust model as an approximation for a fluid with small but non-vanishing speed of sound is  misleading.
\item Rigorous analysis implies that in the region \mbox{$1-\frac{2}{3\alpha}<K<\frac{2}{3}-\frac{1}{3\alpha}$} shock formation is caused by the coupling of the fluid equations and not an isolated effect of the Burgers term in the fluid velocity equation.
\end{itemize}
In addition to these results, we also provide constructive analytical evidence that dust ($K=0$) is unstable in the region $0<\alpha\leq 1/2$, and radiation fluids ($K=1/3$) are unstable for $0<\alpha\leq 1$. While the latter has already been established with different methods in \cite{Sp13}, the former was previously unknown. 

\subsubsection{Methodology}
Using energy methods based on analytical techniques from hyperbolic PDEs, we establish a lower bound for the location of the critical line. This bound coincides precisely with \eqref{transitionline}. We prove that all points $(\alpha, K)$ below that line are stable. In the complementary region of parameter space, we use high precision numerics and corresponding asymptotic analysis to provide strong evidence for shock formation. Both approaches independently identify the critical line \eqref{transitionline} as the location for the phase transition between stability and shock formation.

\section{Equations}
We study the relativistic Euler equations \eqref{Euler} on spacetimes of the form \eqref{spacetimemetric}
with toroidal topology, $g_{ij}=\delta_{ij}$ and impose $\mathbb T^2$-symmetry on the fluid variables, energy density $\rho=\rho(t,x)$ and fluid velocity $u^{\mu}=\left(u^{0}(t,x),u^{1}(t,x),0,0\right)$, where $u_\mu u^\mu=-1$ and $\rho(t,x)=\rho(t,x+2\pi)$ etc.  

	We introduce \emph{expansion-normalized variables} 
	\begin{equation*}
		v:= t^{\alpha}\frac{u^{1}}{u^0}, \quad L:= \log\left(t^{3\alpha(1+K)}\rho\right),
	\end{equation*}
	to obtain the system 
\eq{\label{eq:evol}\alg{
\p t v&=-\frac{\alpha(1-3K)}tv-t^{-\alpha}\frac{1-K}{1-Kv^2}v\p x v\\
&\quad-t^{-\alpha}\frac{K}{1+K}\frac{(1-v^2)^2}{1-Kv^2}\p xL \\
&\quad+ \alpha (1-K)(1-3K)\frac{t^{-\alpha}}{1-Kv^2}\,v^3,\\
\\
\p t L&=-\frac{1+K}{1-Kv^2}t^{-\alpha}\p x v-\frac{1-K}{1-Kv^2}t^{-\alpha}v \p xL\\
&\quad+ \alpha(1+K)(1-3K)\frac{t^{-\alpha}}{1-Kv^2}\,v^2.
}}
Additionally, for the mean velocity $
\overline v=\int_{\mathbb S^1} v dx$ the following equation holds:
\eq{\label{meanvalue}
\p t\overline v=-\frac{\alpha(1-3K)}{t} \overline v+\int_{\mathbb S^1} \frac{v^2t^ {-\alpha}}{1-Kv^2}h(\p x L,v) dx,
}
where $h(\cdot \,,\cdot )$ is a homogeneous polynomial.
\section{Decay of perturbations in the stable regime}
In the stable regime of parameter space, $(1-K)\alpha>2/3$,  the following theorem establishes stability for each point in that region.
For convenience, we introduce suitable Sobolev norms, measuring the perturbation from the quiet fluid:
\eq{\alg{
\|(v,L)\|_{(2)}&:= \|\nabla v\|_{L^2}+\|\nabla^2 v\|_{L^2}\\
&\qquad+\|\nabla L\|_{L^2}+\|\nabla^2 L\|_{L^2}.
}}
\begin{thm}\label{thm-main}
Let $(1-K)\alpha>2/3$, $0<K<1/3$.  Then, for given $\delta>0$ there exists an $\varepsilon>0$ such that 
for initial data $(v_0,L_0)$ with
$|\overline v_0|+\|(v_0,L_0)\|_{(2)}<\varepsilon$
the emanating solution $(v(t),L(t))$ decays according to
\eq{\alg{
\|(v(t),L(t))\|_{(2)}&\lesssim t^{-\alpha(1-3K)/2+\delta}
}}
and
\eq{
|\overline v(t)|\lesssim t^{-\alpha(1-3K)+\delta}.
}
Similar estimates hold for higher order norms.
\end{thm}
The mechanism driving decay, which yields Theorem \ref{thm-main}, is presented below; for a complete proof without symmetry restrictions see \cite{FOW:2024}. Decay is established by deriving an estimate on $\overline v$ and on the $L^2$-energy of derivatives of $v$ and $L$ and combining them using a continuity (or bootstrap) argument.
For that purpose, choose a sufficiently small constant $0<\delta<2/5K$ obeying $10\delta<\alpha(1-3K)-2(1-\alpha)$ and consider a sufficiently small initial data size $\varepsilon>0$ so that the bounds 
\eq{\label{bootstrap}\alg{
&\|(v,L)\|_{(2)}\leq C_1t^{-\alpha(1-3K)/2+\delta}\\
&\mbox{ and }|\overline v|\leq C_2 t^{-\alpha(1-3K)+\delta}
}}
hold on a sufficiently long time interval. We will then show that improved bounds hold for all times. 

\begin{figure*}[ht]
	\includegraphics[width=0.95\columnwidth]{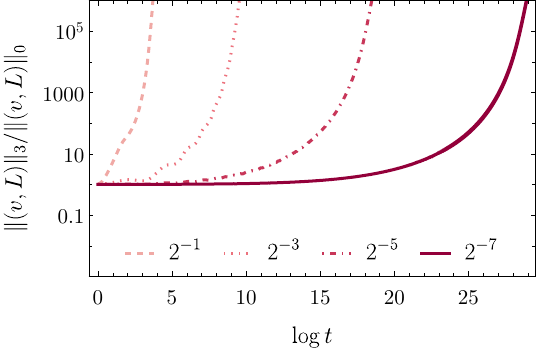}
	\hspace{6ex}
	\includegraphics[width=0.95\columnwidth]{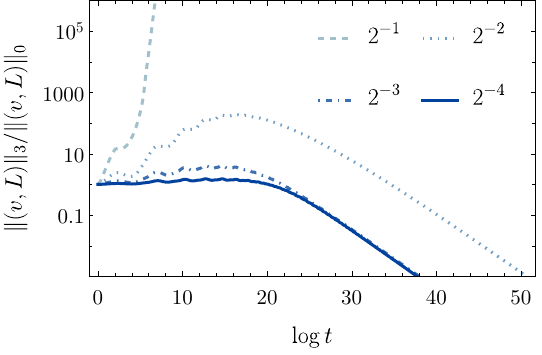}	
	\caption{Time evolution of $\|(v,L) \|_{3}/\|(v,L) \|_{0}$, see Eq.~\eqref{eq:nomrs}, for initial data $(v,L)|_{t=1}=(\varepsilon\sin{x},0)$ with different amplitudes $\varepsilon$ (colour coded). 
	We show both the unstable $(\alpha,K)=(0.7,1/6)$ (left) and stable $(\alpha,K)=(0.9,1/6)$ (right) cases.
	}
\label{fig:n1}
\end{figure*}

From \eqref{meanvalue}, we obtain for the rescaled mean value $\widehat v:=t^{\alpha(1-3K)-\delta/2}\overline v$, the following equation:
\eq{
\frac{d}{dt}\widehat v=-\frac{\delta}{2t}\widehat v+\frac{t^{\alpha(1-3K)-\delta/2+1-\alpha}}{t}f(\overline v,\nabla v,\nabla L),
}
where 
$
|f(\overline v,\nabla v,\nabla L)|\leq F(\overline v,\|( v, L)\|_{(2)})
$
and $F$ is a polynomial, of at least third order, without lower order terms. In consequence, a factor of $t^{3\alpha(1-3K)/2-3\delta}$ can be absorbed into $f$ yielding a  term uniformly bounded by \eqref{bootstrap}. This implies
$$
\frac{d}{dt} {\widehat v}\leq-\delta t^{-1}\widehat v + \frac{C_3}{t^{1+(\alpha(1-3K)-2(1-\alpha))/4}}.
$$
As the exponent of $t$ in the denominator is strictly larger than 1 by the condition on $\alpha$ and $K$, this implies the desired bound for $\overline v$.

We then define the following energy functional measuring the deviation from the quiet fluid state:
\eq{\alg{
E_c[v,L](t)&=\frac12\int_{\mathbb S^1} (\partial_x v)^2 dx\\
&\quad + \frac{K}{(1+K)^2}\frac12\int_{\mathbb S^1} (1-v^2)^2(\partial_x L)^2 dx\\
&\qquad+c t^{-(1-\alpha)}\int_{\mathbb S^1} v\cdot\partial_xL dx
}}
with parameter $c>0$ chosen as $c=\frac12\alpha(1-3K)(1+K)^{-1}$.

For sufficiently large $t$ and sufficiently small $|v|$ this is equivalent to the $L^2$-norm of $\partial_x v$ and $\partial_x L$. To see this, we have $\int v \partial_x L dx= \int (v-\bar v) \partial_x L dx$, using integration by parts, and the latter can be estimated up to a constant by the product $\|\partial v\|_{L^2}\cdot\|\partial L\|_{L^2}$ using H\"older's estimate and the Poincar\'e inequality.
For the corrected energy of second order $E^{(2)}_c(t)=E_c[v(t),L(t)]+E_c[\partial_x v(t),\partial_xL(t)]$, we obtain an estimate of the form
\eq{\alg{
\frac{d}{dt} E^{(2)}_c(t)&\leq -\frac{\alpha (1-3K)+C_2 t^{-(1-\alpha)}}{t}E^{(2)}_0(t) \\
&+C_1 \frac{t^{1-\alpha}}t \left[ \left(E^{(2)}_0\right)^{3/2}
+N\left(\sqrt{E_0^{(2)}},\overline v\right)\right],
}}
where $N$ is a polynomial, of at least third order, without lower order terms. The condition $\alpha(1-3K)>2(1-\alpha)$ allows a factor $t^{\alpha(1-3K)-\delta/2}$ to be absorbed into the energy to obtain
\eq{
\frac{d}{dt} \left(t^{\alpha(1-3K)-\delta/2}E^{(2)}_c[v,L](t)\right)\leq C\frac{t^{1-\alpha}}{t} t^{\alpha(1-3K)-\delta/2}\overline v^3
}
for sufficiently small initial data. By the condition on $\alpha$ and $K$, the right-hand side is integrable and this implies, for sufficiently small data, that the energy decays as
\eq{
E^{(2)}_c[v,L](t)\leq \frac{C_4}{t^{\alpha(1-3K)-\delta/2}}.
}
This concludes the proof of the theorem.

Under the assumption that the contribution of the derivative of the energy density is small a similar analysis as above would imply that the fluid stabilizes in the region $3K<2-1/\alpha$. This follows since the correction term would not be necessary and in turn the full decay provided by the negative definite term in $v$ would be available to compensate the $t^\alpha$ growth. However, the numerics show instability in the region
$
1-\frac{2}{3\alpha}<K<\frac23-\frac{1}{3\alpha},
$
in particular. In consequence, the shocks forming in that region cannot be generated solely by the Burgers term in the fluid velocity equation but must involve an effect caused, at least in part, by the derivative of the energy density.

\section{Numerics on small data shock formation} 
We next solve \eqref{eq:evol} numerically with the method of lines: we use a Fourier pseudospectral approach to discretise the equations in the $x$-direction on a grid of $N$ collocation points $x_{j}=2\pi j/N$, $j=0,\ldots,N-1$, combined with an adaptive Runge-Kutta algorithm of variable order and step-size to evolve the resulting equations forward in time. For convenience, we work with logarithmic time $\tau=\ln t$. In the time-integration we set a strict error tolerance bound to $10^{-14}$. For spatial resolution, we typically take $N=128$, which is as tradeoff between accuracy and the runtime. We stress that for a fixed number of grid points we can follow the singularity formation (in the unstable regime) for a finite amount of time. As a consequence, we are unable to get an exact breaking time $ t_{*}$, as that would require using an adaptive grid. However, by carefully choosing $N$ and the thresholds in the shock formation/stability detection algorithm (see below), we obtain robust and consistent results. 

We experimented with several families of initial data. Our classification of the parameter space $(\alpha, K)$ is independent of these choices. For concreteness, we discuss the results with $(v,L)|_{t=1}=(\varepsilon\sin{x},0)$, $\varepsilon>0$.

During the time-evolution we monitor the quantity $\|(v,L) \|_{3}/\|(v,L) \|_{0}$ where
\begin{equation}
	\label{eq:nomrs}
	\|(v,L) \|_{k} := (1+K)^{2}\|\nabla^{k}v\|^{2}_{L^{2}} + K\|\nabla^{k}L\|^{2}_{L^{2}}
	\,,
\end{equation}
and we classify a solution as smooth or shock forming if this ratio is smaller than $10^{-3}$ or larger than $10^{6}$ respectively. See Fig.~\ref{fig:n1}, which illustrates a behaviour of solutions in the limit of $\varepsilon\rightarrow 0$ for different points in the parameter space $(\alpha,K)$.

\begin{figure*}[th]
	\includegraphics[width=0.95\columnwidth]{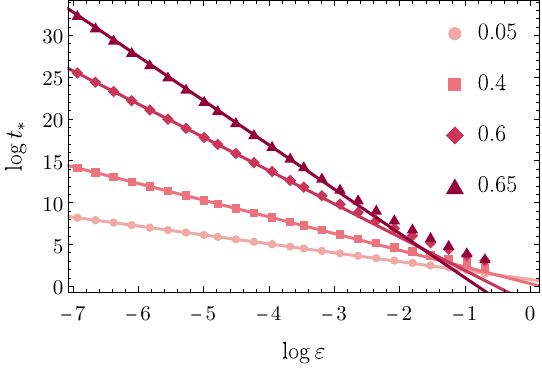}
	\hspace{6ex}
	\includegraphics[width=0.95\columnwidth]{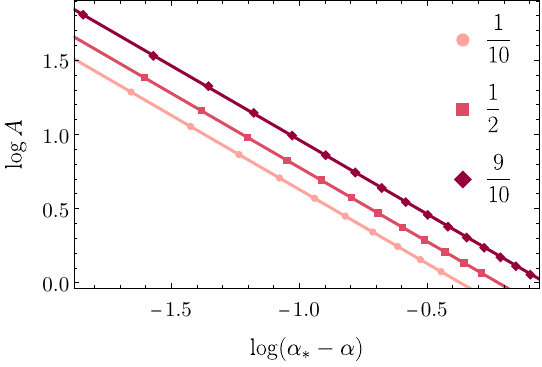}
	
	\caption{
	Left: The logarithm of the breaking time  $t_{*}$ as a function of initial data amplitude $\varepsilon$ for $K=1/6$ and sample values of $\alpha$ (colour coded) left to the transition curve. Points are the numerical data and the lines are the fits of \eqref{eq:scaling1} in the asymptotic region $\varepsilon\rightarrow 0$.
	Right: The behaviour of the proportionality coefficient $A$ in \eqref{eq:scaling1} as a function of the expansion exponent $\alpha$. On the log-log plot the data points, shown are the results for a few values of $K$ (colour coded), lay on the straight line, implying the relation \eqref{eq:scaling2}, with $\alpha_{*}$ values which agree with Eq.~\eqref{transitionline}.
	}
\label{fig:n3n4}
\end{figure*}

We consider individual points in the parameter space $(\alpha,K)$ from within the regime of decelerated expansion rates and speeds of sound below radiation. A direct determination of the boundary between the stable and unstable regions appears particularly difficult as the time evolution in the transition region is especially demanding. Therefore, we rely on the following asymptotic analysis.

We start the analysis from the points with very slow expansion and move towards larger $\alpha$ along the $K=const$ lines. In the unstable regime we observe that the breaking time exhibits the scaling
\begin{equation}
	\label{eq:scaling1}
	\log t_{*} \approx -A \log \varepsilon + B\,, \quad \varepsilon\rightarrow 0
	\,,
\end{equation}
with $(\alpha,K)$-dependent constants $A$ and $B$, which can be determined by fitting the formula to the numerical data, see Fig.~\ref{fig:n3n4}. Next, inspecting the relation between $A$ and $\alpha$, with fixed $K$, we find
\begin{equation}
	\label{eq:scaling2}
	A \approx \frac{D}{\alpha_* - \alpha}
	\,,
\end{equation}
ie.~$A$ diverges as $\alpha$ approaches $\alpha_{*}$ from below. As before the constants $D$ and $\alpha_{*}$ can be found by fitting the formula, cf.~Fig.~\ref{fig:n3n4}. Such values of $\alpha_*$ all lie on the critical line, see Fig.~\ref{fig:n3n4}. The relative error between $\alpha_{*}$ and the value from Eq.~\eqref{transitionline} is less than $0.8\%$, and it systematically decreases when we increase both the resolution $N$ and the threshold for shock formation and simultaneously get closer to the critical line. This agrees with the analytical results of previous sections, from the side of the unstable region. For $D$ in \eqref{eq:scaling2}, we have the following guess $D\approx K+2/3$. Together \eqref{eq:scaling1} and \eqref{eq:scaling2} imply the following for the breaking time
\begin{equation}
t_{*} \sim \varepsilon^{\left(K+2/3\right)/\left(\frac{2}{3(1-K)}-\alpha\right)}
	\,,
	\quad \alpha<\frac{2}{3(1-K)}\,.
\end{equation}

For evolution on the critical line $(\alpha,K_{\mathrm{crit}})$, we find evidence that
\begin{equation}
	t_{*} \sim \exp\left(e^{B}/\varepsilon\right)
	\,,
	\quad
	B=const
	\,,
	\quad \alpha=\frac{2}{3(1-K)}\,,
\end{equation}
which implies instability as $\varepsilon\rightarrow 0$.
The points right of (or below) the critical line are classified as stable, since for them the breaking times $t_{*}\rightarrow\infty$ for $\varepsilon\rightarrow \varepsilon_0>0$, and the solution exists globally for $\varepsilon<\varepsilon_{0}$.

The unstable region matches the rigorously known boundary values on the radiation line $K=1/3$ (cf.~following section). The vertical rightmost boundary of the stable region ($\alpha$=1) consists of stable points if $K<1/3$,  and one unstable point for $K=1/3$, which connects with the unstable region and hence consists of a transition point (both stable and unstable points are in its neighborhood). The lower horizontal boundary of the stable region coincides only partially with the dust line for $\alpha>2/3$. This suggests that the transition from the unstable region to the dust line for $1/2<\alpha<2/3$ is not continuous.

\section{Bounds on the shock formation time}
In the cases of dust ($K=0$) and radiation ($K=1/3$) we provide rigorous estimates on the breaking time using intersecting characteristics. This implies a definitive proof of the \emph{instability of quiet dust solutions} on spacetimes of the form  \eqref{spacetimemetric} with $\alpha\leq 1/2$ as well as of the \emph{instability of quiet radiation solutions} with $\alpha\leq1$. The estimates are derived by constructing explicit intersecting curves, which envelop two characteristics. 

The equation for the fluid velocity of a dust field in a spacetime of the form \eqref{spacetimemetric} is, due to decoupling, given by 
\eq{\label{dust}
\p tu+\left(\sqrt{t^{2\alpha}u^2+1}\right)^{-1}u \p xu=-2\alpha t^{-1} u.
}
We consider two distinct points $x_1,x_2\in[0,2\pi]$ with $x_1<x_2$ and $u_0(x_1)>u_0(x_2)>0$.
The characteristic emanating from the point $x_i$ is
\eq{
\dot x^i(t)=\frac{u_0(x_i)t^{-2\alpha}}{\sqrt{1+t^{-2\alpha}u_0(x_i)^2}}.
}
In the case $\alpha =1/2$, we define the curves
$
\gamma_i^{+}(t)=u_0(x_i)\log t+x_i
$
for $i=1,2$. Then $\gamma_2^{+}$ is an upper bound to $x^2(t)$ when $t>1$. 
A lower bound to $x^1(t)$ is given by
$\gamma_{1}^-=\gamma^+_1-m(u_0(x_1)),$
%
where
$$
m(y) = \left|2y\log\left(2y^{-2}\left(-1+\sqrt{y^{2}+1}\right)\right)\right|.
$$
The enveloping curves $ \gamma_{1}^{-} $ and $ \gamma_{2}^{+} $ intersect at time
\eq{\label{BT-onehalf-dust}
T_+=\exp\left(-\frac{x_2-x_1}{\Delta(u_0)}\right)\exp\left(\frac{m(u_0(x_1))}{-\Delta(u_0)}\right),
}
where $\Delta(u_0)=u_0(x_2)-u_0(x_1)$. The breaking time $ T_{+} $ is finite, proving instability of the solution for all initial data in question. Furthermore, by rescaling a particular initial data profile by $ \lambda$, one can deduce that the exponential term tends to $ 1 $ as $ \lambda\to0 $. This gives a particularly simple expression for the estimate of the breaking time. 
For $0<\alpha<1/2$ similar upper and lower bounds $ \gamma^{+} $ and $ \gamma_{-} $ can be established (for details see \cite{FOW:2024}). 

In the case of radiation, we choose $\alpha\in (0,1]$ and define $u\in (-1,1)$ via $u^0=(1-u^2)^{-1/2}$ and $u^1=u(1-u^2)^{-1/2}$. Following \cite{ReSt08} we define
\eq{
\varphi=\frac{\sqrt{K}}{1+K}\ln (\rho/\rho_c),
}
where $\rho_c>0$ is a constant and
\eq{\alg{
r=\varphi+\frac12\ln\frac{1+u}{1-u},\quad s=\varphi-\frac12\ln\frac{1+u}{1-u}.
}}
Finally, $\mathbf R:=t^{-c}e^r$ and $\mathbf S:=t^{-c}e^s$ for some fixed $ c>0 $. Along characteristics it can be shown that $\mathbf R$ is constant and if $\mathbf S$ is chosen constant initially then it remains constant globally. For the first spatial derivative of $\mathbf R$, $\mathbf W=\partial_x \mathbf R$, we obtain the Riccati-type equation (for details see \cite{FOW:2024})
\eq{\label{Riccati}
\frac{d\mathbf W}{dt}= \frac{12 \mathbf S(3\alpha)^{1-\alpha}}{\left[(3+\sqrt 3)\mathbf R+(3-\sqrt 3)\mathbf S\right]^2}t^{-\alpha}\mathbf W^2.
}
In combination with the behaviour of $\mathbf R$ and $\mathbf S$, this equation leads to blow-up in finite-time of non-trivial initial data for $\mathbf W$, ie.~the spatial derivative of $ \mathbf{R} $, for all $\alpha\leq 1$.

\section{Discussion}

We provide strong evidence for a phase transition of barotropic fluids in decelerated spacetimes and classify all points in the decelerated region of the parameter space with respect to the (in)stability nature of the corresponding barotropic fluid. In combination with previous results on the regime of accelerated and linear
expansion, this resolves the fluid stabilization problem for barotropic fluids and provides a first generic example of a phase transition from stable to unstable cosmological fluids. The analytic and numerical approaches are complementary and predict the critical line independently of each other. 

While this class of fluids is the most relevant in cosmology, there are numerous alternative equations of state used in modelling other regimes of relativistic matter dynamics. One could also look at non-compact spatial topologies where the effect of dispersion would also occur, see for example \cite{Taylor} in the context of Vlasov matter. Given the robust nature of the methods employed in the study of our barotropic case, we expect that similar effects exist for alternative equations of state and for other topologies. There is no a priori reason why the critical behaviour should be limited to the barotropic case. 


\begin{acknowledgements}
\begin{center}\textbf{Acknowledgements}\end{center}
D.F.~and M.O.~acknowledge support by the Austrian Science Fund (FWF) \href{http://doi.org/10.55776/P34313}{P 34313-N}. M.O.~acknowledges financial support by the Vienna Doctoral School in Physics (VDSP). The research of M.M.~was supported by the Austrian Science Fund (FWF) through Project \href{http://doi.org/10.55776/P36455}{P 36455} and the START-Project \href{http://doi.org/10.55776/Y963}{Y 963}.
\end{acknowledgements}

\end{document}